\documentclass[preprint,prd]{revtex4}

\usepackage{amsbsy}
\usepackage{amssymb}
\usepackage[sumlimits]{amsmath}
\usepackage{graphicx}
\usepackage[active]{srcltx}
\usepackage{pdfsync}

\usepackage{color}

\usepackage{cancel}

\def\x{{\mathrm{x}}}
\def\y{{\mathrm{y}}}

\def\n{{\mathrm{n}}}
\def\b{{\mathrm{b}}}

\def\s{{\mathrm{s}}}

\def\n{{\rm n}}
\def\p{{\rm p}}
\def\e{{\rm e}}

\def\c{{\rm c}}

\def\s{\mathrm{s}}

\def\n{{\rm n}}
\def\p{{\rm p}}
\def\e{{\rm e}}

\def\e{{\rm e}}
\def\s{{\rm s}}
\def\c{{\rm c}}

\def\be{\begin{equation}}
\def\ee{\end{equation}}
\def\beq{\begin{equation}}
\def\eeq{\end{equation}}
\def\bea{\begin{eqnarray}}
\def\eea{\end{eqnarray}}

\def\bear{\begin{eqnarray}}
\def\eear{\end{eqnarray}}

\begin{document}

\title{Beyond ideal magnetohydrodynamics: From fibration to 3+1 foliation}

\author{N. Andersson$^1$, I. Hawke$^1$, K. Dionysopoulou$^1$ and G.L. Comer$^2$}

\affiliation{
$^1$ Mathematical Sciences and STAG Research Centre, University of Southampton,
Southampton SO17 1BJ, United Kingdom\\
$^2$ Department of Physics, Saint Louis University, St. Louis, MO, 63156-0907, USA}

\begin{abstract}
We consider a resistive multi-fluid framework from the 3+1 space-time foliation point-of-view, paying particular attention to issues relating to the use of multi-parameter equations of state and the associated inversion from evolved to primitive variables. We highlight relevant numerical issues that arise for general systems with relative flows.  As an application of the new formulation, we consider a three-component system relevant for hot neutron stars. In this case we let the baryons (neutrons and protons) move together, but allow heat and electrons to exhibit relative flow. This reduces the problem to three momentum equations; overall energy-momentum conservation, a generalised Ohm's law and a heat equation. Our results provide a hierarchy of increasingly complex models and prepare the ground for new state-of-the-art simulations of relevant scenarios in relativistic astrophysics.
\end{abstract}

\maketitle

\section{Context}

A range of astrophysical phenomena involve violent nonlinear matter dynamics. The modelling of such systems requires fully nonlinear multi-dimensional simulations taking into account the live spacetime of general relativity. In recent years there has been considerable progress in developing the required computational tools, especially for archetypal gravitational-wave sources like supernova core collapse~\cite{SNreview} and neutron star mergers~\cite{baiottirezzollareview}. The technology is now reaching the point where the consideration of more sophisticated matter models is required. In the case of supernova modelling, it is well known that the neutrinos play an important role in triggering the explosion itself~\cite{Janka2012} and the role of magnetic fields may also be significant~\cite{moestanature}. For neutron star mergers, finite temperature effects are central as shock heating ramps up the temperature of the merged object to levels beyond that expected even during core collapse (see, e.g., \cite{Bauswein2010} or \cite{Kastaun2015}). Dynamical magnetic fields are likely to have decisive impact on the post-merger dynamics and  may leave an observational signature, e.g.\ in short gamma-ray bursts (e.g., \cite{Kumar2015}).

To suggest that consistent modelling of the required physics is challenging would be an understatement. Hence, it is natural that progress has been made by adding individual ingredients one by one. However, this strategy can be problematic as there may be an interplay between the different physics aspects. With this in mind, it makes sense to consider the formulation of a new generation of models which include the key physics from the outset. This should allow us to identify (and quantify the relevance of) any issues that may be overlooked in current simulations. It should also enable progress towards (even) more sophisticated simulations, once the computational technology makes such work feasible.

The problems we want to investigate have the common feature that they involve the flow of a number of identifiable ``currents'' beyond that of the bulk matter flow associated with a perfect fluid. In the first instance, we have the charge current associated with electromagnetism, at finite temperature heat will flow and for mature neutron stars there may also be a relative flow associated with the presence of superfluid components. As full kinetic simulations of these kinds of systems pose enormous challenges, it is natural to take as a starting point the well-developed framework for relativistic multi-fluid dynamics~\cite{livrev,monster}. We have already considered the fundamental aspects of the problem~\cite{variate} and the connection with the involved microphysics and the features that arise in models of increasing complexity~\cite{fibrate}. In the latter case we introduced a fibration of spacetime associated with a specific set of fluid observers. This approach is natural if one is mainly interested in the local fluid dynamics (e.g.~wave propagation) and it also leads to the 1+3 formulation often used in cosmology (where ``clocks'' associated with the fluid observers define the notion of cosmic time), see~\cite{tsagas} for a relevant discussion. This approach is, however, not natural for  nonlinear  simulations with a live spacetime. Instead, most such work makes use of a 3+1 spacetime foliation (see~\cite{baum} for a relevant discussion), where progression towards the ``future'' is associated with a set of Eulerian observers. Hence, it is relevant to complement the discussion in~\cite{fibrate} by extending the multifluid model from fibration to foliation.

The aim of this paper is to develop the 3+1 version of the general framework discussed in~\cite{fibrate}. The main aspects remain the same -- in particular, we introduce a set of fluid observers to make contact with thermodynamics and the microphysics associated with the equation of state -- but the foliation approach leads to new issues that need to be resolved (e.g.~the inversion from evolved to primitive variables).
In order to keep the discussion tractable, we focus on a three-component system relevant for hot neutron stars. We assume neutrons remain non-superfluid and locked to the protons, but let heat and electrons exhibit relative flow.  In effect, this reduces the problem to three momentum equations; overall energy-momentum conservation, a generalised Ohm's law and a heat equation. Our formulation of these equations should allow us to build models with causal heat flow~\cite{heat1,heat2} and non-ideal magnetohydrodynamics features associated with resistive scattering~\cite{namhd}.

The models we consider in this paper do not account for neutrinos, the emission of which will have significant impact on the evolution of a hot system, or the elastic neutron star crust, which will be relevant for mature (cold) systems. Both these aspects can be accounted for in the general formalism. In fact, a formulation for simulating elastic models was recently presented in~\cite{elastic} and this model extends directly to our framework. When it comes to the neutrinos, the hot models we develop here may in principle contain trapped neutrinos (forming part of the entropy component) but we do not account for radiative fluxes. Standard approaches for including relativistic radiation transport, such as~\cite{Cardall2013} or \cite{Shibata2014} could be employed to extend the model, but we leave this for future work.

Before we proceed it is also worth making a comment on notation. We distinguish between three sets of indices. We use  $a,b,c,...$ for spacetime indices and $i,j,k,...$ for spatial indices on each spatial slice. These indices satisfy the Einstein summation convention, as usual. We also use indices $\x,\y,...$ to label the different fluid components. The summation convention does not apply to these indices.

\section{3+1 basics}

Following the standard approach to formulate the equations of motion in a way suitable for numerical simulations (see, e.g.~\cite{Alcubierre:2008}), we foliate spacetime into a family of spacelike hypersurfaces $\Sigma_t$ which arise as level surfaces of a scalar time $t$. Given the normal to this surface
\be
N_a = - \alpha \nabla_ a t \ , 
\label{normal}
\ee
we have
\be
N_a = (-\alpha,0,0,0) \ ,
\ee
and the normalisation $N_a N^a=-1$ leads to $\alpha^2 = -1/g^{tt}$.
The sign in \eqref{normal}  ensures that time flows into the future. The function $\alpha$ is known as the lapse. The dual to $\nabla_a t$ leads to a time vector
\be
t^a = \alpha N^a + \beta^a \ ,
\ee
where the so-called shift vector $\beta^a$ is spatial, which means that $N_a \beta^a = 0$. It follows that
\be
N^a = \alpha^{-1} ( 1,-\beta^i) \ ,
\ee
and the spacetime can be written in the standard ADM form:
\be
ds^2 = - \alpha^2 dt^2 + \gamma_{ij} \left( dx^i + \beta^i dt \right)   \left( dx^j + \beta^j dt \right) \ ,
\ee
where the (induced) metric on the spacelike hypersurface is
\be
\gamma_{ab} = g_{ab} + N_a N_b \ .
\ee
We note that $\gamma^a_b$ represents the projection orthogonal to $N_a$ and that $\gamma_{ab}$ and its inverse can be used to raise and lower indices of purely spatial tensors. For example, we have $\beta_i = \gamma_{ij} \beta^j$.

In essence, the lapse $\alpha$ determines the rate at which proper time advances from one time slice to the next, along the normal  $N_a$, and the shift vector $\beta^i$ determines how the coordinates  shift from one spatial slice to the next. The two functions encode the coordinate freedom of general relativity.

Reading off the metric from the line element, we have
\be
g_{ab} = \left(\begin{array}{cc} -\alpha^2 + \beta_i \beta^i & \beta_i \\ \beta_i & \gamma_{ij} \end{array} \right) \ ,
\ee
with inverse
\be
g^{ab} = \left(\begin{array}{cc} -1/\alpha^2 & \beta^i/\alpha^2 \\ \beta^i/\alpha^2 & \gamma^{ij}-\beta^i \beta^j /\alpha^2 \end{array} \right) \ .
\ee

Given the spacetime foliation, we can decompose any tensor quantity into time and space components. For example,
let us assume that we have a fluid associated with a four velocity $u^a$. Then we can introduce the decomposition~\footnote{We are using the convention that all velocities measured by the Eulerian observer have hats, while the velocities relative to the fluid frame do not.}
\be
u^a = W (N^a + \hat v^a) \ ,
\label{fluidobs}
\ee
where $N_a \hat v^a =0$ and the Lorentz factor is given by
\be
W= - N_a u^a =  \alpha u^t = (1-\hat v_i \hat v^i)^{-1/2} \ ,
\ee
(the last equality follows from $u^a u_a=-1$).
From this, it is easy to see that
\be
\hat v^t = 0 \ , \qquad \hat v^i = {u^i\over W} - N^i = {1\over \alpha} \left( {u^i \over u^t} + \beta^i\right) \ ,
\ee
and it follows that
\be
\hat v_t = g_{ta}v^a = \beta_i \hat v^i \ , \qquad \hat v_i = \gamma_{ia} \hat v^a = {\gamma_{ij} \over \alpha} \left( {u^j\over u^t} + \beta^j \right) \ .
\ee

Finally, we need to consider derivatives. First of all, we need a derivative associated with the hypersurface. Thus we introduce the (totally) projected derivative
\be
D_a = \gamma_a^b \nabla_b \ ,
\ee
where all free indices should be projected into the surface.
This derivative is compatible with the spatial metric in the sense that
\be
D_a\gamma_{bc} = \gamma_a^d \gamma_b^e \gamma_c^f\nabla_d \gamma_{ef} = 0 \ ,
\ee
which means that it acts as a covariant derivative in the surface orthogonal to $N^a$. Hence, it is straightforward to construct a tensor algebra for the three-dimensional spatial slices.
In particular, we can introduce a three-dimensional Riemann tensor. This projected Riemann tensor obviously does not contain all the information from its four-dimensional counterpart. The missing information is encoded in the extrinsic curvature, $K_{ab}$. This is a symmetric spatial tensor, such that $N^a K_{ab}=0$, which measures (roughly speaking) how the $\Sigma_t$ surfaces  curve relative to the spacetime. In practice, we measure how the normal $N_a$ changes as it is parallel transported along the hypersurface. That is,  we define
\be
K_{ac} = -D_a N_c = - \gamma_a^b \gamma_c^d \nabla_b N_d = - \nabla_a N_c - N_a (N^b\nabla_b N_c) \ ,
\label{Kdef}
\ee
where the second term is an analogue of the fluid four-acceleration. We also have
\be
K= K^a_a = g^{ab}K_{ab} = \gamma^{ab} \gamma_{ab} = - \nabla_a N^a \ .
\ee
Alternatively, we can use the properties of the Lie derivative to show that
\be
K_{ij} = - 2\mathcal L_N \gamma_{ij} \ ,
\ee
but since
\be
\mathcal L_N = {1\over \alpha} ( \mathcal L_t - \mathcal L_\beta) = {1\over \alpha } ( \partial_t - \mathcal L_\beta) \ ,
\label{tder}
\ee
we have
\be
\partial_t \gamma_{ij} = - 2\alpha K_{ij} + \mathcal L_\beta \gamma_{ij} \ .
\ee
From the trace of this expression we get
\be
\alpha K = - \partial_t \ln \gamma^{1/2} + D_i \beta^i \ ,
\ee
where $\gamma=g^{ab}\gamma_{ab}$ and $\gamma^{ij} \partial_t \gamma_{ij} = \partial_t \ln \gamma$.

\section{Perfect fluids}

The (standard) results in the previous section provide the tools we need to make progress in deriving the 3+1 version of relativistic fluid dynamics and/or the Einstein field equations (the interested reader can find useful reviews of the spacetime problem in~\cite{Alcubierre:2008} or \cite{baumgarte2010numerical}). Our main interest here is the equations of fluid dynamics. We want to develop a version of the multi-fluid models outlined in~\cite{fibrate} suitable for numerical evolutions. As this systems builds on -- and extends -- the simple perfect fluid model, it is natural to start by reviewing the standard approach (see~\cite{Fontlivrev} for more details).

\subsection{Baryon number conservation}

 Let us start with the simple case of baryon number conservation. That is, we assume the flux $n u^a$ is conserved, where $n$ is the number density according to an observer moving along with the fluid. Thus we have
\be
\nabla_a (n u^a) = \nabla_a [ Wn (N^a + \hat v^a) ]= 0 \ .
\ee
First we note that the particle number density measured by the Eulerian observer is
\be
\hat n =-N_a n u^a = nW \ ,
\ee
so we have
\be
N^a \nabla_a \hat n + \nabla_i (\hat n \hat v^i) = - \hat n \nabla_a N^a =  \hat n K \ ,
\ee
(since $\hat v^i$ is spatial). Making use of the Lie derivative and \eqref{tder} we have
\be
N^a \nabla_a \hat n = \mathcal L_N \hat n = {1\over \alpha} ( \partial_t - \mathcal L_\beta) \hat n = - \nabla_i (\hat n \hat v^i) + \hat n K
\ ,
\ee
or
\be
\partial_t \hat n + (\alpha \hat v^i - \beta^i )\nabla_i \hat n + \alpha \hat n \nabla_i \hat v^i = \alpha  \hat n K \ .
\ee
Finally, since $\hat v^i$ and $\beta^i$ are already spatial, we have
\be
\partial_t \hat n + (\alpha \hat v^i - \beta^i )D_i \hat n + \alpha \hat n D_i \hat v^i = \alpha  \hat n K =- \hat n   \partial_t \ln \gamma^{1/2} + \hat n D_i \beta^i \ ,
\ee
or
\be
\partial_t \left( \gamma^{1/2} \hat n\right) +  D_i \left[ \gamma^{1/2}\hat n (\alpha \hat v^i - \beta^i )\right] = 0 \ ,
\label{baryons}
\ee
where we have used the fact that
\be
\left( -g\right)^{1/2} = \alpha \gamma^{1/2} \ ,
\ee
so
\be
 \nabla_a (-g)^{1/2} = \nabla_a ( \alpha \gamma^{1/2}) = 0 \ .
\ee
For future reference, it is also worth noting that $D_i \gamma =0$, so we have
\be
D_i \gamma^{1/2} = \partial_i \gamma^{1/2} - \Gamma^j_{ji} \gamma^{1/2} \ ,
\ee
where the Christoffel symbol is the one associated with the covariant derivative in the hypersurface.

The final result, \eqref{baryons}, simply represents the advection of the baryons along the flow, as seen by the (fixed) Eulerian observer.

\subsection{The energy/momentum equations}

Moving on, the fluid equations of motion follow from $\nabla_a T_\mathrm{M}^{ab}=0$ where the standard
 case of a perfect fluid (ignoring electromagnetism for the moment) is described by the stress-energy tensor
\be
T_\mathrm{M}^{ab} = (p+\varepsilon) u^a u^b + p g^{ab} \ .
\ee
Here $p$ and $\varepsilon$ are the pressure and the energy density, respectively. As discussed in~\cite{fibrate} these quantities are related by the equation of state, which encodes the relevant microphysics. In order to make contact with the underlying physics, a numerical simulation must allow the extraction of these quantities.

A numerical simulation is naturally carried out using quantities measured by the Eulerian observer. That is, we decompose the stress-energy tensor into normal and spatial parts as
\be
T_\mathrm{M}^{ab} = \rho N^a N^b + 2 N^{(a} S^{b)} + S^{ab} \ ,
\label{TM}
\ee
with
\be
 \rho = N_a N_b T^{ab} =  \varepsilon W^2 - p \left( 1 - W^2\right) \ ,
\ee 
\be
S^i = - \gamma^i_c N_d T^{cd} =  \left(p+\varepsilon \right) W^2 \hat v^i \ ,
\ee
and
\be
S^{ij} = \gamma^i_c \gamma^j_d T^{cd} = p \gamma^{ij}  + \left( p +\varepsilon \right) W^2  \hat v^i \hat v^j \ .
\ee

A projection of the equations of motion along $N_a$ then leads to the energy equation. From
\be
N^a \nabla_a   \rho +   \rho \nabla_a N^a + \nabla_ a S^a - N_b N^a \nabla_a S^b - N_b \nabla_a S^{ab} =   0 \ ,
\ee
we get
\be
N^a \nabla_a  \rho + \nabla_a S^a=  \rho K-S^b N^a\nabla_a N_b - S^{ab}\nabla_a N_b \ ,
\ee
and
\be
{1\over \alpha} \left( \partial_t - \mathcal L_\beta\right)   \rho + \nabla_a S^a=  \rho K-S^b D_b \ln \alpha  + S^{ab}K_{ab} \ .
\ee
Finally, we arrive at
\be
\partial_t  \left(\gamma^{1/2}  \rho\right) + D_i \left[ \gamma^{1/2} \left( \alpha S^i -\rho \beta^i\right) \right]  =  \gamma^{1/2} \left( \alpha S^{ij}K_{ij} -S^i D_i  \alpha 
\right) \ .
\label{energyq}
\ee

Note that, it is common to evolve $\tau = \rho - m_0 \hat n$ (where $m_0$ is the baryon rest mass density) rather than $\rho$. This is done to avoid numerical issues arising from the fact that \eqref{energyq}
matches (to leading order in velocity) the evolution equation for the conserved proper rest-mass density [$m_0$ times \eqref{baryons}]. This change has no impact on the formal discussion in the rest of this paper, but it is important to keep it in mind, nevertheless.

Note also that, one may opt to evolve the entropy instead of the energy~\cite{fibrate}. A basic Newtonian calculation (see, e.g., \cite{Leveque2002}) shows that the energy equation leads directly to an advection equation for the entropy. However, the energy equation is typically preferred in numerical work as its balance law form is compatible with standard conservative schemes and ensures suitable behaviour when shocks appear. The equivalence between the two formulations breaks down for more complex systems (with additional components), leading to questions as to which description is more natural. We will touch on this issue when we discuss the inversion from evolved to primitive variables for  multifluid systems in Section VF.

Turning to the momentum equation, which is obtained by a projection orthogonal to $N_a$, we have
\be
 \rho N^a \nabla_a N^c + \gamma^c_{\ b}N^a \nabla_a S^b + S^c \nabla_a N^a + S^a \nabla_a N^c + \gamma^c_{\ b} \nabla_a S^{ab}
=0 \ ,
\ee
which leads to
\be
\left( \partial_t - \mathcal L_\beta\right) S_i - S^j \left( \partial_t - \mathcal L_\beta\right)\gamma_{ij} - \alpha K S_i + \rho D_i \alpha
+ \alpha \gamma_{ij} D_k S^{kj} = 0 \ ,
\ee
where we have used
\be
N^a \nabla_a S^c = \mathcal L_N S^c + S^a \nabla_a N^c = \mathcal L_N S^c - S^a K_a^c \ .
\ee
This leads to the final result
\be
\partial_t (\gamma^{1/2} S_i) + D_j \left[ \gamma^{1/2} \left( \alpha S_i^j -S_i \beta^j \right) \right] = \gamma^{1/2} \left( S_j D_i \beta^j - \rho D_i \alpha \right) \ .
\label{momentum}
\ee

\subsection{Conservative to primitive}

We now have the set of evolution equations we need for the fluid part of the single-component problem. However, one important issue remains to be resolved. We need to consider the inversion from the variables obtained from the evolution to the primitive fluid variables associated with the equation of state. We need to understand this issue because it highlights the link to the underlying microphysics and we will need to generalise this strategy later when we consider more complex settings.

Let us, for simplicity, consider the case of a cold barotropic fluid, such that the equation of state provides the energy as a function of the baryon number density $\varepsilon = \varepsilon(n)$. This then leads to the chemical potential
\be
\mu = {d\varepsilon \over dn} \ ,
\ee
and the pressure $p$ follows from the thermodynamic relation:
\be
p = n \mu - \varepsilon \ .
\label{pdef}
\ee
Basically, in order to connect with the thermodynamics, we need the evolved number density. We also need to ``decide'' which observer ``measures'' equation of state quantities. In the single-fluid case the second question is relatively easy to answer; we need to express the equation of state in the co-moving fluid frame (associated with $u^a$). In the multi-fluid case, the answer is not as straightforward.

In the barotropic case, the  evolution system \eqref{baryons} and \eqref{momentum} provides (assuming that $\gamma^{1/2}$ is known from the evolution of the Einstein equations)
\be
\hat n = nW = n (1-\hat v^2)^{-1/2} \ ,
\label{one}
\ee
and
\be
S^i = (p+\varepsilon) W^2 \hat v^i \ .
\label{two}
\ee
We need to invert these two relations to get the primitive variables $n$ and $\hat v^i$. This can be  formulated as a one-dimensional root-finding problem. For example, we could guess $n=\bar n$. This then allows us to work out $\varepsilon$ from the equation of state and $p$ from \eqref{pdef}. With these variables in hand we can solve
\be
{S^2 \over (p+ \varepsilon)^2} =  W^4 \hat v^2 \ ,
\ee
for $\hat v^2$.  This allows us to work out the Lorentz factor $W$ and then $\hat v^i$ follows from \eqref{two}. Finally, we get
$n=\hat n/W$ from \eqref{one}. The result  can be compared to our initial guess $\bar n$. Iterating the procedure gives a  solution consistent with the conserved quantities, and hence all primitive quantities.

This procedure is straightforward but it is easy to see that the inversion may be much more involved for more
 complex problems. In fact, the problem is tricky already at the level of standard ideal magnetohydrodynamics. As this is an important issue for the extended models we aim to develop it is worth explaining the issue in  detail.

In general,
the electromagnetic dynamics is fully specified in terms of  the vector potential $A^a$, but as in~\cite{fibrate} it may be more intuitive to work with the electric and magnetic fields.  In the 3+1 decomposition, where the observer is associated with $N^a$, we then have  the Faraday tensor
\be
F_{ab} =  2 N_{[a} E_{b]}  + \epsilon_{abcd}N^c B^d \ .
\label{Faraday}\ee
That is, the electric and magnetic fields measured in the Eulerian frame are
\be
E_a = -  N^b F_{ba} \ ,
\ee
and
\be
B_a = - N^b \left( {1 \over 2} \epsilon_{abcd}F^{cd}\right) \ .
\ee
The  fields are both orthogonal to $N^a$, so
each  has three components, just as in non-relativistic physics.

In order to account for the electromagnetic contribution to the stress-energy tensor (see Appendix) we need
\be
T^\mathrm{EM}_{ab} = {1\over \mu_0} \left[ g^{cd} F_{ac} F_{bd} - {1\over 4} g_{ab} (F_{cd} F^{cd} ) \right] \ .
\ee
In terms of the fields (measured by the Eulerian observer) we have
\be
T^\mathrm{EM}_{ab} = E^2 N_a N_b + E_a E_b + \gamma_{ab} B^2 - B_a B_b + 2N_{(a}\epsilon_{b)dh} e^d B^h - {1\over 2} g_{ab} \left( B^2 - E^2\right) \ ,
\ee
where we have introduced
\be
\epsilon_{abc} = \epsilon_{dabc} N^d  \ .
\ee
This means that the total stress-energy tensor takes the form \eqref{TM}, with
\be
 \rho = \varepsilon W^2 - p \left( 1 - W^2\right) + {1\over 2\mu_0} \left( E^2 + B^2\right) \ ,
 \label{magrho}
\ee
\be
S^i  =  \left(p+\varepsilon \right) W^2 \hat v^i  + {1\over \mu_0}  \epsilon^{ijk}E_j B_k \ ,
\label{magmom}
\ee
and
\be
S^{ij} = p \gamma^{ij}  + \left( p +\varepsilon \right) W^2  \hat v^i  \hat v^j - {1\over \mu_0} \left[ E^i E^j + B^i B^j - {1\over 2} \left( E^2 + B^2\right) \gamma^{ij} \right] \ .
\label{magsij}
\ee

From these expressions we learn that,
when electromagnetism is added, we either have no conceptual change to the inversion strategy or things get considerably more complicated. The conserved fluid variables remain the number density $\hat n$, the momenta $S^i$ (now defined in \eqref{magmom} and still evolved by \eqref{momentum}) and the energy $\rho$ (now defined in \eqref{magrho} and evolved by \eqref{energyq}). In addition, we have the electric and magnetic fields, which are evolved by the usual Maxwell equations (see Appendix).

Now, if we retain both electric and magnetic fields in the evolution then a direct algebraic calculation takes us from the magnetised energy in \eqref{magrho} and the momentum in \eqref{magmom} to their  fluid counterparts. Hence, we can still use the one-dimensional root finding strategy from the pure fluid problem,

However, in ideal magnetohydrodynamics, the electric field is not evolved, but  computed from a constraint. This reduces the number of evolution equations and ensures that, for example, the ``${\bf E} = - {\bf v} \times {\bf B}$'' constraint holds identically. The constraint relating electric and magnetic fields requires the velocity, which is one of the primitive variables we need to compute. This considerably complicates the inversion process (see~\cite{mart}, section 5.8 for a discussion of the various options used in the literature).

\section{Adding degrees of freedom}

Building on the discussion in~\cite{fibrate}, let us now consider the multifluid aspects of the problem. We will divide the discussion into two parts. In this first section, we consider general aspects without committing ourselves to a specific model (or choice of fluid frame). In the next section, we make the analysis problem specific by focussing on the equations that are required to model the dynamics of  hot magnetised neutron stars.

\subsection{Non-conserved fluxes}

In a general multifluid problem, we have a number of distinct fluxes $n_\x^a= n_\x u_\x^a$, where the x labels each fluid. These fluxes are not necessarily conserved, so we have
\be
\nabla_a n_\x^a = \Gamma_\x \ ,
\ee
where $\Gamma_\x$ is the relevant reaction rate. In the 3+1 formulation, we need
\be
n_\x^a = n_\x W_\x ( N^a + \hat v_\x^a) =  \hat n_\x ( N^a + \hat v^a_\x) \ ,
\ee
where $\hat n_\x$ is the  number density measured by the Eulerian observer, $\hat v_\x^a$ is the corresponding fluid velocity and
\be
W_\x = (1- \hat v_\x^2)^{-1/2} \ ,
\ee
is the Lorentz factor.

We now have
\be
\nabla_a \left( \hat n_\x N^a +  \hat n_\x \hat v_\x^a\right)  = \Gamma_\x \ ,
\ee
which leads to [following the steps that led to \eqref{baryons}]
\be
\partial_t \left( \gamma^{1/2} \hat n_\x \right) + D_i \left[ \gamma^{1/2} \hat n_\x (\alpha \hat v_\x^i - \beta^i) \right] = \alpha \gamma^{1/2} \Gamma_\x \ .
\label{nex}\ee
This is (again) an advection equation, but it also allows the model to account for possible nuclear reactions. In the following, we will work with the number densities $\hat n_\x$, but it is worth noting that it would be straightforward to replace these with particle fractions $x_\x = \hat n_\x/ \hat n$ (once we have a definition of the ``total'' number density -- see e.g.\ section VA) should one want to do so.

\subsection{Individual momentum equations}

In the multifluid model, the equations that represent total energy and momentum conservation are replaced (or complemented, see~\cite{fibrate} for a discussion) by a set of individual momentum equations. If we allow for particle reactions and resistivity, these take the form~\cite{variate} 
\be
2 n_\x^b \nabla_{[b}\tilde \mu^\x_{a]}  + \tilde \mu^\x_a \Gamma_\x
= R^\x_a  \ ,
\label{intmom}\ee
or
\be
2 n_\x^b \nabla_{[b}\mu^\x_{a]}  + \mu^\x_a \Gamma_\x
= j_\x^b F_{ab} + R^\x_a - e_\x \Gamma_\x A_a \ ,
\ee
where the canonical momentum is
\be
\tilde \mu^\x_a = \mu^\x_a + e_\x A_a \ ,
\ee
with $e_\x$ the charge per particle of the x-fluid and $A_a$ the electromagnetic vector potential. The gauge issues associated with the explicit presence of the vector potential have been discussed in~\cite{variate}. The equation of motion~\eqref{intmom} has the hydrodynamical forces (including the Lorentz force) and the ``rocket'' term associated with particle creation on the left hand side balancing the resistivity on the right hand side.

As discussed in~\cite{variate}, a general model takes the form
\be
R^\x_a = \tilde \mu_\x \Gamma_\x u^\x_a   + \sum_{\y\neq\x} \perp^b_{\x a} \mathcal R^{\x\y}_b \ ,
\ee
with $\tilde \mu_\x = -u_\x^a \tilde \mu^\x_a$. In order to be more specific, we make use of the phenomenological model from~\cite{variate}. This involves introducing relative flows with respect to a chosen fluid observer (with four velocity $u^a$), such that
\be
u_\x^a = \gamma_\x (u^a + v_\x^a)  \qquad \mbox{with} \qquad \gamma_\x = (1-v_\x^2)^{-1/2} \ ,
\label{fluid2}
\ee
(where the fluid frame Lorentz factor $\gamma_\x$ is not to be confused with $\gamma=\gamma^i_{\ i}$ for the space-time).
The resistivity is then given by
\be
R^\x_a =  \Gamma_\x \tilde \mu_\x u^\x_a +  \sum_{\y\neq\x} \mathcal R^{\x\y} (\delta_a^b + v_\x^b u_a)  w^{\y\x}_b \ ,
\label{pheno}
\ee
where $w^{\y\x}_b = v^\y_b - v^\x_b$ is the velocity difference, for all material particles. The construction is then closed by the constraint on the resistivity that enters the entropy equation ($\x=\s$)
\be
R^\s_a = - \sum_{\x\neq\s} R^\x_a \ ,
\ee
which means that (recalling that $T=\mu_\s$)
\be
{1\over \gamma_\s} T \Gamma_\s = - (u^a + v_\s^a) R^\s_a = (u^a + v_\s^a) \sum_{\x\neq\s} R^\x_a
= - \sum_{\x\neq\s} \sum_{\y\neq\x} \mathcal R^{\x\y} w_{\x\s}^a w^{\y\x}_b \ge 0 \ ,
\label{TGs}
\ee
and the $\mathcal R^{\x\y}$ coefficients are required to be positive by the second law of thermodynamics (they are also symmetric in $\x$ and $\y$).

\subsection{The 3+1 form of the momentum equations}

Let us now return to \eqref{intmom}. In order to work out the spatial component of this equation, we  need the explicit form of the conjugate momentum. Hence, we make the decomposition
\be
\mu_a^\x = \hat \mu_\x N_a + S^\x_a \ ,
\ee
which introduces the chemical potential according to the Eulerian observer, $\hat \mu_\x$, and where the flux $S^\x_a$ may account for entrainment (as we will explain later).

In general, we need (for each fluid component)
\be
\gamma^a_c \left( 2 n_\x^b \nabla_{[b}\mu^\x_{a]} + \mu^\x_a \Gamma_\x \right) = \mathcal F^\x_c \ ,
\label{fluideq}
\ee
where
\be
 \mathcal F^\x_c = \gamma^a_c \left[  j_\x^b F_{ab} + R^\x_a - e_\x A_a \Gamma_\x  \right] \ .
 \ee

Leaving the right-hand side of \eqref{fluideq} aside for the moment, we have
\begin{multline}
\gamma^a_c \left( 2 n_\x^b \nabla_{[b}\mu^\x_{a]} + \mu^\x_a \Gamma_\x \right) = {1 \over \alpha} \left[ (\partial_t - \mathcal L_\beta) \left( \hat n_\x S^\x_c \right)  + \hat n_\x D_c (\alpha \hat \mu_\x)  \right] + D_b \left( \hat n_\x \hat v^b_\x S^\x_c \right) \\
- \hat n_\x \hat v^b_\x D_c S^\x_b + \hat n_\x S_\x^b K_{bc} -  \hat n_\x K S^\x_c \ ,
 \end{multline}
and final equation takes the form
\begin{multline}
(\partial_t - \mathcal L_\beta) \left( \hat n_\x S^\x_c \right)  + \hat n_\x D_c (\alpha \hat \mu_\x)   + \alpha D_b \left( \hat n_\x \hat v^b_\x S^\x_c \right)
- \alpha \hat n_\x \hat v^b_\x D_c S^\x_b \\
= \alpha \mathcal F^\x_c -\alpha  \hat n_\x S_\x^b K_{bc} +\alpha S^\x_c\hat n_\x K \ , 
\end{multline}
or\begin{multline}
\partial_t ( \gamma^{1/2} \hat n_\x S^\x_i ) + D_j \left[  \gamma^{1/2}  \hat n_\x \left(\alpha \hat v^j_\x -  {\beta^j  } \right) S^\x_i  \right] + \hat n_\x D_i \left( \alpha \gamma^{1/2} \hat \mu_\x\right) -  \hat n_\x \hat v^j_\x D_i \left( \alpha \gamma^{1/2} S^\x_j \right) \\
= \gamma^{1/2} \left[ \alpha \mathcal F^\x_i - \alpha \hat n_\x S_\x^j K_{ij} +  \hat n_\x S^\x_j D_i\beta^j \right] \ .
\label{finaleq}
\end{multline}

Let us now consider the right-hand side. We need
\be
\mathcal F^\x_c
 = \gamma^a_c \left[  j_\x^b F_{ab} +  \Gamma_\x ( \tilde \mu_\x u^\x_a - e_\x A_a) +  \sum_{\y\neq\x} \mathcal R^{\x\y} (\delta_a^b + v_\x^b u_a)  w^{\y\x}_b  \right] \ ,
 \label{rhseq}
\ee
where
\be
\gamma^a_c  j_\x^b F_{ab} = e_\x \hat n_\x  \left( E_c +     \epsilon_{cbd} \hat v_\x^b B^b \right) \ ,
\ee
and
\be
\gamma^a_c ( \tilde \mu_\x u_a^\x - e_\x A_a) = \hat \mu_\x \hat v^\x_a +  [ \gamma^a_c + W_\x^2 \hat v_c^\x ( N^a + \hat v^a_\x) ] A_a \ .
\ee

In order to work out the final term, we need to consider the microphysics. This is naturally done in the (suitably defined) ``fluid'' frame~\cite{fibrate}. From \eqref{fluidobs} and \eqref{fluid2} it follows that
\be
v_\x^a = - \left( W - {W_\x \over \gamma_\x} \right) N^a - W \hat v^a + {W_\x \over \gamma_\x} \hat v_\x^a \ ,
\ee
such that
\be
W_\x = \gamma_\x(1-N_a v_\x^a) \ ,
\ee
and
\be
\gamma_\x = W W_\x ( 1- \hat v_\x^a \hat v_a) \ .
\label{Wxeq}
\ee
This last result is important because all quantities on the right-hand side are evaluated in the Eulerian frame, and can be (at least in principle) extracted from the evolution.

After a bit of algebra, we find that
\be
\gamma^a_c  (\delta_a^b + v_\x^b u_a)  w^{\y\x}_b  = {W_\y \over \gamma_\y} \hat v^\y_c - {W_\x \over \gamma_\x} \hat v^\x_c
+ W \hat v_\c \left[ {1\over \gamma_\x^2} - {W_\y \over \gamma_\y} {W_\x \over \gamma_\x} (1 - \hat v_\x^b \hat v^\y_b) \right] \ ,
\ee
where, given \eqref{Wxeq}, all quantities on the right-hand side can be expressed in terms of Eulerian quantities.

\subsection{The total momentum equation}

As discussed in~\cite{fibrate} the single-fluid equations discussed in Section~III will, in general, take a different form in the multi-fluid case.

In particular, in the multifluid case the stress-energy tensor takes the form
\be
T_\mathrm{M}^{ab} = \Psi g^{ab} + \sum_\x n_\x^a \mu_\x^b \ ,
\ee
where
\be
\Psi = \Lambda - \sum_\x n_\x^a \mu^\x_a \ .
\ee
In terms of the Eulerian observer we have
\be
T_\mathrm{M}^{ab} = \Psi g^{ab} + \sum_\x \hat n_\x (N^a+\hat v_\x^a) ( \hat \mu_\x N^b + S_\x^b) \ .
\ee
In the general case, which accounts for entrainment between different flowing components~\cite{livrev}, we have
\be
\mu_\x^a  = \mathcal B^\x n_\x^a + \sum_{\y\neq\x} \mathcal A^{\x\y} n_\y^a \ ,
\ee
such that
\be
\mu_\x^a  = \mathcal B^\x \hat n_\x (N^a + \hat v_\x^a)  + \sum_{\y\neq\x} \mathcal A^{\x\y} \hat n_\y (N^a + \hat v_\y^a) \ .
\ee
Thus we see that
\be
\hat \mu_\x=  \mathcal B^\x \hat n_\x + \sum_{\y\neq\x} \mathcal A^{\x\y} \hat n_\y \ ,
\ee
and
\be
S_\x^a =  \mathcal B^\x \hat n_\x \hat v_\x^a + \sum_{\y\neq\x} \mathcal A^{\x\y} \hat n_\y  \hat v_\y^a
=  \hat \mu_\x \hat v_\x^a + \sum_{\y\neq\x} \mathcal A^{\x\y} \hat n_\y  ( \hat v_\y^a - \hat v_\x^a) \ .
\ee
When we add the contributions to the stress-energy tensor, we see that
\be
T_\mathrm{M}^{ab} = \Psi g^{ab} + \sum_\x \hat n_\x \hat \mu_\x (N^a+\hat v_\x^a) (N^b+\hat v_\x^b) +
\sum_\x \sum_{\y\neq\x} \hat n_\x \hat n_\y \mathcal A^{\x\y} ( N^a + \hat v_\x^a) \hat w_{\y\x}^b \ .
\ee

Comparing to \eqref{TM} we find that we now need
\be
 \rho = N_a N_b T_\mathrm{M}^{ab} = \sum_\x \hat n_\x \hat \mu_\x -\Psi \ ,
 \label{energysum}
\ee
\be
S^a = - \gamma^a_c N_d T_\mathrm{M}^{cd} = \sum_\x \hat n_\x \left( \hat \mu_\x  - \sum_\x \sum_{\y\neq\x} \hat n_\x \hat n_\y \mathcal A^{\x\y} \right) \hat v_\x^a = \sum_\x \hat n_\x^2 \mathcal B^\x \hat v_\x^a \ ,
\ee
and
\begin{multline}
S^{ab} = \gamma^a_c \gamma^b_d T_\mathrm{M}^{cd} =  \Psi \gamma^{ab} + \sum_\x \hat n_\x   \hat v_\x^a \left( \hat \mu_\x\hat v_\x^b + \sum_{\y\neq\x} \hat n_\y \mathcal A^{\x\y} \hat w_{\y\x}^b \right) \\
= \Psi \gamma^{ab} + \sum_\x \hat n_\x^2  \mathcal B^\x  \hat v_\x^a \hat v_\x^b +
\sum_\x  \sum_{\y\neq\x} \hat n_\x  \hat n_\y \mathcal A^{\x\y} \hat v_\x^a \hat v_\y^b \ .
\end{multline}

In principle, the multi-fluid model is now complete and we can turn our attention to the physics.
However, the  complexity of the problem means that it is sensible to consider a specific setting and it is also wise to introduce  simplifications. Hence, we will focus on developing a model relevant for hot magnetised neutron stars, where the electrons flow relative to the baryons (neutron and protons) and where the dynamics of the thermal component is retained.

\section{Application: Hot magnetised stars}

Let us consider the specific problem of hot neutron stars (above the critical temperature for superfluidity).  We then have the equations for baryon number conservation and total momentum conservation from before. Once we account for heat- and charge currents, we have a three-component problem. We need a system of equations for the baryon number density $\hat n$ and the (Eulerian) fluid velocity $ \hat v^i$, the electron number  density $\hat n_\e$ and the charge current $\hat J^i$, and the entropy density $\hat s$ and the heat flux  $Q^i$.  That is, we are dealing with a problem with three distinct fluxes. The purpose of this section is to define the relevant quantities,  derive the equations that govern them and devise a strategy that allows the inversion from evolved variables to the primitive variables used to describe the microphysics.

\subsection{Baryon number conservation}

It is natural to begin by revisiting the issue of baryon number conservation. In the general multi-fluid case, where neutrons and protons are not locked together, we still need to impose
\be
\Gamma_\n + \Gamma_\p = 0 \ .
\ee
This means that we can add the individual continuity equations to get
\be
\partial_t \left[ \gamma^{1/2} (\hat n_\n+  \hat n_\p)  \right] +  D_i \left[ \alpha \gamma^{1/2} \left(  \hat n_\n \hat v_\n^i  +  \hat n_\p  \hat v_\p^i \right)- \gamma^{1/2}(\hat  n_\n + \hat n_\p ) \beta^i \right]= 0\ .
\ee
The baryon number measured by the Eulerian observer is
\be
\hat n = \hat n_\n+  \hat n_\p \ ,
\ee
and we see that we retain the standard single-fluid result provided that we introduce
\be
\hat n \hat v^i =  \hat n_\n \hat v_\n^i  +  \hat n_\p  \hat v_\p^i \ .
\label{frame}
\ee
This is tantamount to working in a fluid frame analogous to the Eckart frame familiar from considerations of relativistic heat flux (see~\cite{fibrate,heat1} for discussion). If we work in a different frame, which we are perfectly free to do, then the baryon conservation law will necessarily be different.

Given the central role that the baryon number density plays in the problem, we will assume that $\hat v^i$ is defined by \eqref{frame} in the following.  This means that baryon number conservation is ensured by
\be
\partial_t \left[ \gamma^{1/2}\hat n \right] +  D_i \left[ \gamma^{1/2}  \hat n \left(  \alpha    \hat v^i-  \beta^i \right) \right]= 0 \ ,
\ee
as usual.

We arrive at the same conclusion by assuming that the neutrons and protons are locked (e.g. assuming effective interparticle scattering, leading to a short relative mean-free path) such that
\be
\hat v^i = \hat v_\n^i = \hat v_\p^i \ .
\ee
This assumption would have  been sufficient for the present discussion,  but it is useful to know that the result holds more generally.

\subsection{Momentum conservation}

Next, we need the equations for the energy and the total momentum. The energy $\rho$ is given by \eqref{energysum} and evolved by \eqref{energyq}.  If we ignore entrainment  (the main mechanism for which is anyway due to a relative drift between neutrons and protons) then  the total fluid contribution to the (Eulerian) momentum flux is
\begin{multline}
S^i =  \sum_\x \hat n_\x \hat \mu_\x \hat v_\x^i = (\hat n_\n \hat \mu_\n + \hat n_\p \hat \mu_\p) \hat v^i + \hat n_\e \hat \mu_\e \hat v_\e^i + \hat s \hat T \hat v_s^i \\
= (\rho + \Psi)  \hat v^i +  \hat n_\e \hat \mu_\e ( \hat v_\e^i -  \hat v^i) +  \hat s \hat T ( \hat v_\s^i -  \hat v^i) \ ,
\label{momdef}
\end{multline}
where $\hat n_\s = \hat s$ and $\hat \mu_\s = \hat T$ is the temperature measured by an Eulerian observer. The relevant evolution equation is (still) \eqref{momentum}.

Later we will find it more convenient to replace the electron velocity with the charge current and the entropy velocity with the heat flux.
We first of all need the charge current
\be
j^a = e (n_\p^a -n_\e^a)=  e ( \hat n_\p - \hat n_\e)  N^a + e  (  \hat n_\p  \hat v^a - \hat n_\e \hat v_\e^a)  = \hat \sigma N^a + \hat J^a \ ,
\ee
with $N_a \hat J^a = 0$.
From this  we see that
\be
 \hat \sigma =  e ( \hat n_\p - \hat n_\e) \ ,
\ee
and
\be
 \hat J^a = e  ( \hat n_\p  \hat v^a - \hat n_\e \hat v_\e^a) \longrightarrow \hat v_\e^a = {1\over \hat n_\e} \left( \hat n_\p  \hat v^a - {\hat J^a \over e}\right) \ .
\ee
Next, introduce the heat flux (relative to the fluid frame) as
\be
Q^i = {\hat s \hat T}  ( \hat v_\s^i - \hat v^i) \ .
\ee

In terms of these new variables, we  have
\be
S^i =   (\rho + \Psi)  \hat v^i +  { \hat \mu_\e \over e} \left( \hat \sigma \hat v^i - \hat J^i \right) +    Q^i \ .
\ee

Similarly, we get
\begin{multline}
S^{ij} = \Psi \gamma^{ij} + \sum_\x \hat n_\x \hat \mu_\x  \hat v_\x^i \hat v_\x^j \\
= \Psi \gamma^{ij} + \left[\rho+\Psi + \left(\hat n_\p^2-\hat n_\e^2\right){\hat \mu_\e \over n_\e} \right] \hat v^i \hat v^j\\
 - 2{\hat n_\p \hat \mu_\e \over e \hat n_\e^2} \hat v^{(i}\hat J^{j)} + 2  \hat v^{(i} Q^{j)} + {\hat \mu_\e \over e^2 \hat n_\e} \hat J^i \hat J^j + {1\over \hat s \hat T} Q^i Q^j \ .
\end{multline}

\subsection{A linear drift model}

As discussed in~\cite{fibrate} it is natural to assume that the drift velocities in the fluid frame are small, such that $v_\x^a\ll1$ and $\gamma_\x\approx 1$. This should be a realistic assumption for many physical situations. In essence, this assumption allows us to linearise the problem in the relative fluxes which simplifies the problem considerably and makes the connection with the microphysics encoded in the equation of state more straightforward.

If the relative drift of each fluid is small in the frame associated with $\hat v^i$, then the difference between $\hat v^i$ and $\hat v_\x^i$ must be small, as well. Retaining only linear terms we have
\be
v_\x^i = W [ \delta^i_j +W^2 \hat v_j (N^i+\hat v^i)] (\hat v_\x^j - \hat v^j) \ ,
\ee
which means that
\be
w_{\y\x}^i = W [ \delta^i_j +W^2 \hat v_j (N^i+\hat v^i)] \hat w_{\y\x}^j \ ,
\ee
and the resistivity \eqref{pheno} simplifies dramatically. We now have
\be
\gamma^a_c  (\delta_a^b + v_\x^b u_a)  w^{\y\x}_b  \approx W [ \delta^b_c + \hat v_c \hat v^b W^2] \hat w^{\y\x}_b \ .
\ee
We also need
\be
W_\x \approx W [ 1 + W^2 \hat v_c(\hat v_\x^c -\hat v^c)] \ .
\ee

In addition to linearising in the drift velocities, it makes sense to assume that the system is charge neutral in the fluid frame. We then have $n_\e = n_\p$ and it follows that
\be
\hat \sigma \approx e n_\e W^3 ( \hat v_j \hat w_{\p\e}^j) \ .
\ee
We also have
\be
\hat J^i \approx e n_\e W (\delta^i_j +W^2 \hat v^i \hat v_j) \hat w_{\p\e}^j \ ,
\ee
which leads to
\be
\hat v_\e^i \approx \hat v^i - {1\over e n_\e W} (\delta^i_j - \hat v_j \hat v^i) \hat J^j \ ,
\ee
and we see that $W_\e \approx W$.
It also follows that
\be
\hat \sigma \approx  \hat v_j  \hat J^j \ ,
\ee
which makes intuitive sense.

Similarly, for the entropy component we have
\be
\hat v_\s^i \approx \hat v^i + {1\over sT W^2} Q^i \ .
\ee

In order to close the system, we need a multiparameter equation of state.
In the fluid frame, we (quite generally, as long as we ignore entrainment) have an equation of state of form $\varepsilon = \varepsilon(n_\x)$, such that, in the case of small drift velocities;
\be
\varepsilon = u_a u_b T^{ab}_\mathrm{M} = - \Psi + \sum_\x n_\x \mu_\x \ .
\ee
From this we see that the local pressure is $p=\Psi$ and we have
\be
p +\varepsilon = \sum_\x n_\x \mu_\x \ .
\ee
Moreover, the individual chemical potentials follow from
\be
\mu_\x = \left( {\partial \varepsilon \over \partial n_\x} \right)_{n_\y} \ , \qquad \y \neq \x \ .
\ee
That is, at this level of approximation, we retain the familiar thermodynamical relations and in the case we are considering we need an equation of state of form $\varepsilon = \varepsilon(n, n_\e, s)$.

It follows that
\be
p + \rho \approx (p+ \varepsilon) W^2 - 2W^2 \hat v_i Q^i \ .
\ee
We also have
\be
S^i \approx   (\rho + p) \hat v^i +  {\mu_\e W \over e} \left( \hat \sigma \hat v^i - \hat J^i \right) +    Q^i \ ,
\ee
and
\be
S^{ij} \approx p \gamma^{ij} + \left[\rho+p + {1\over e} \mu_\e W \hat \sigma \right] \hat v^i \hat v^j  - {2\mu_\e W \over e} \hat v^{(i} \hat J^{j)}
+ 2\hat  v^{(i} Q^{j)} \ .
\ee

\subsection{Ohm's law}

In the multifluid model, Ohm's law follows from the electron momentum equation~\cite{fibrate}.  
Using $\x=\e$ in \eqref{finaleq} we get
\begin{multline}
\partial_t ( \gamma^{1/2} \hat n_\e S^\e_i ) + D_j \left[  \alpha \gamma^{1/2}  \hat n_\e \left( \hat v^j_\e - {\beta^j \over \alpha } \right) S^\e_i  \right] + \hat n_\e D_i \left( \alpha \gamma^{1/2} \hat \mu_\e\right) -  \hat n_\e \hat v^j_\e D_i \left( \alpha \gamma^{1/2} S^\e_j \right) \\
= \gamma^{1/2} \left[ \alpha \mathcal F^\e_i - \alpha \hat n_\e S_\e^j K_{ij} +  \hat n_\e S^\e_j D_i\beta^j \right] \ ,
\label{finohm}
\end{multline}
where $\mathcal F^\e_i$ follows from \eqref{rhseq}.

As we are ignoring entrainment we have
 \be
S^\e_i = \hat \mu_\e  \hat v_\e^i \approx \mu_\e \left( W \hat v^i - {\hat J^i\over e n_\e
}\right)  \ .
\ee
Making use of this in \eqref{finohm} (and linearising in the relative fluxes) we arrive at the final momentum equation for the charge current.

In the following we  will ignore particle reactions. That is, we take $\Gamma_\e=0$, which has the benefit of removing electromagnetic gauge issues from the problem (as the explicit dependence on the vector potential is gone).

With these assumptions we have
\begin{multline}
\mathcal F^\e_i \approx  -  e \hat n_\e  \left( E_i +     \epsilon_{ijk} \hat v_\e^j B^k \right)
+ W( \delta^j_i + \hat v_i \hat v^j W^2) \sum_{\y\neq \e} \mathcal R^{\e\y}  \hat w^{\y\e}_j \\
\approx (e n_\e W - \hat \sigma) E_i + \epsilon_{ijk} ( e n_\e W \hat v^j - \hat J^j) B^k \\
+ {1\over e n_\e} ( \mathcal R^{\e\b} + \mathcal R^{\e\s}) \hat J_i + {1\over sTW}  \mathcal R^{\e\s} ( \delta^j_i + \hat v_i \hat v^j W^2) Q_j \ .
\end{multline}

In order to evolve the equation for the charge current, we need the electron number density (or some proxy for it). At the level of approximation we are working, it follows from \eqref{nex} that
\be
\partial_t (\gamma^{1/2} \hat n_\e ) + D_i \left\{ \gamma^{1/2} \left[ \hat n_\e ( \alpha \hat v^i - \beta^i) - {\alpha \over e} ( \hat J^i - \hat \sigma \hat v^i) \right] \right\} = 0 \ .
\ee

\subsection{Heat equation}

In order to account for the flow of heat we need the entropy component.  In this case, it is useful to introduce $s = n_\s$, $s^a = s u_\s^a$ and $T=\mu_\s$ (as before) such that
\be
n_\s^a = s^a = \hat s (N^a + \hat v_\s^a) \ .
\ee
The entropy equation
\be
\nabla_a s^a = \Gamma_\s  \ge 0 \ ,
\ee
then leads to
\be
\partial_t \left( \gamma^{1/2} \hat s \right)  + D_i \left[ \gamma^{1/2} \hat s \left( \alpha \hat v_\s^i - \beta^i\right)\right] = \alpha \gamma^{1/2} \Gamma_\s \ ,
\ee
or, in terms of the heat flux,
\be
\partial_t \left( \gamma^{1/2} \hat s \right) +  D_i \left\{\alpha \gamma^{1/2} \left[{Q^i\over \hat T} +  \hat s \left( \hat v^i- {\beta^i\over \alpha} \right) \right] \right\}= \alpha \gamma^{1/2} \Gamma_\s  \ ,
\label{entrops}
\ee
where $\hat T = - N^a \mu^\s_a$.

Let us now consider the momentum equation \eqref{finaleq} for the thermal component. We need
\be
\mu_a^\s  = \hat  T N_a + S^\s_a  \ .
\ee
From \eqref{finaleq} we then have
\begin{multline}
\partial_t ( \gamma^{1/2} \hat s S^\s_i ) + D_j \left[  \alpha \gamma^{1/2}  \hat s \left( \hat v^j_\s - {\beta^j \over \alpha } \right) S^\s_i  \right] + \hat s D_i \left( \alpha \gamma^{1/2} \hat T\right) -  \hat s \hat v^j_\s D_i \left( \alpha \gamma^{1/2} S^\s_j \right) \\
= \gamma^{1/2} \left[ \alpha \mathcal F^\s_i - \alpha \hat s S_\s^j K_{ij} +  \hat s S^\s_j D_i\beta^j \right]  \ ,
\end{multline}
or
\begin{multline}
\partial_t ( \gamma^{1/2} \hat s S^\s_i ) + D_j \left\{  \alpha \gamma^{1/2}   \left[{Q^j \over \hat T} +\hat s \left(\hat v^j - {\beta^j \over \alpha } \right)\right] S^\s_i  \right\} \\
+ \hat s D_i \left( \alpha \gamma^{1/2} \hat T\right) -  \left( {Q^j \over \hat T} + \hat s \hat v^j\right) D_i \left( \alpha \gamma^{1/2} S^\s_j \right) \\
= \gamma^{1/2} \left[ \alpha \mathcal F^\s_i - \alpha \hat s S_\s^j K_{ij} +  \hat s S^\s_j D_i\beta^j \right]  \ ,
\label{finheat}
\end{multline}
where (as long as  we ignore entropy entrainment)
\be
S_\s^i = \hat T \hat v_\s^i  \approx TW \left( 1 + {1\over sT}\hat v_j Q^j \right) v^i + {1\over sW} Q^i  \ .
\ee
Inserting this in \eqref{finheat} (and linearising in the relative fluxes) we arrive at the final momentum equation for the thermal component.

We also have
\be
\mathcal F^\s_i = \gamma_i^j R^\s_j = -  \gamma_i^j \sum_{\x\neq\s} R^\x_j  \ ,
\ee
which means that
\be
\mathcal F^\s_i = \gamma_i^j R^\s_j \approx -  {1\over e n_\e}  \mathcal R^{\e\s} \hat J_i - {1\over sTW}  ( \mathcal R^{\b\s} + \mathcal R^{\e\s} )( \delta^j_i + \hat v_i \hat  v^j W^2) Q_j  \ .
\ee

Finally, we need an explicit expression for $\Gamma_\s$. We know from \eqref{TGs} that the result will be quadratic in the (fluid frame) drift velocities. Explicitly we have (retaining quadratic terms in the fluxes since they are leading order). 
\begin{multline}
T \Gamma_\s = \mathcal R^{\b\e} w_{\b\e}^2 +  \mathcal R^{\b\s} w_{\b\s}^2 +  \mathcal R^{\e\s} w_{\e\s}^2 \\
\approx  {1\over s^2 T^2 W^2} (\mathcal R^{\b\s}  + \mathcal R^{\e\s}) \left[ Q^2 + (\hat v_j Q^j )^2 W^2 \right]  \\
+ { 2  \over e n_\e sTW} \mathcal R^{\e\s} \hat J^l Q_l  + {1\over e^2 n_\e^2} ( \mathcal R^{\e\s} + \mathcal R^{\e\b} )  ( \hat J^2 - \hat \sigma^2) \ .
\end{multline}

\subsection{Inferring the primitive variables}
\label{sec:c2p_hot}

As discussed in Section~IIIC,  the framework is not complete unless we provide a prescription for working out  the primitive variables from the evolved ones. In the general case, we expect to need to evolve both electric and magnetic fields (or any equivalent set of variables giving the complete Faraday tensor). Thus, unless we make specific simplifications to the model, and as long as we can ignore  gauge issues, we should always be able to calculate all electromagnetic quantities that appear in the evolved variables. This means that when considering the inversion process from conserved to primitive variables, we only need explore the hydrodynamic problem.

In the general case, we have six evolved quantities: We have three scalars:
\begin{eqnarray}
\hat n &=& n W \ , \\
\hat s &=& sW\left( 1 + {1\over sT} \hat v_j Q^j\right) \ ,\\
\hat n_\e &=& n_\e W \ ,
\end{eqnarray}
where we could opt to use the energy $\rho$ instead of $\hat s$, and  three fluxes:
\begin{eqnarray}
S^i &=& \left[ ( p+\varepsilon)W^2 + {\mu_\e \over e} W \hat \sigma - 2W^2 (\hat v_j Q^j) \right] \hat v^i -{\mu_\e W\over e}  \hat J^i + Q^i  \ , \\
S^i_\e &=& \mu_\e \left( W \hat v^i - {1 \over e n_\e} \hat J^i \right)  \ , \\
S^i_\s &=& TW \left( 1 + {1\over sT}\hat  v_j Q^j \right)\hat v^i + {1\over sW} Q^i \ .
\end{eqnarray}

The general problem we have formulated takes us several steps beyond the current state of the art for numerical simulations. However, it is quite easy to strip the model down to a hierarchy of levels. As a first step, let us consider the simple case of a hot single fluid. This is a useful model problem because it illustrates the fact that we may adopt different strategies. If we assume that the entropy is locked to the material component, then we are dealing with a single flow, but we still require a two-parameter equation of state. The usual approach to this problem considers the energy
$\varepsilon$ as the second thermodynamic variable. In this ``energy representation'' the equation of state is, effectively, taken to be of form $p=p(n,\varepsilon)$. The evolution provides values for $\hat n$, $\rho$ and $S_i$. In order to invert this system to the primitive variables we can initiate a root search from a guess $p=\bar p$. By combining the definitions for $\rho$ and $S^i$ we see that
\be
p+ \rho = {S^2 W^2 \over \hat v^2}  \ .
\ee
Given this, we can obtain the Lorentz factor $W$ from the evolved variables and our initial guess $\bar p$. The definition of $\rho$ then provides us with the corresponding value for $\varepsilon$ and the evolved $\hat n$ gives $n$. Now we can work out $p(n,\varepsilon)$ from the equation of state and compare to our guessed value. Iteration of the procedure leads to the solution we need.

The energy approach is straightforward to implement but the multifluid formulation suggests that we may want to consider an alternative approach~\footnote{In principle, the two descriptions are related by the inversion $s=s(n,\varepsilon)$, but this is unlikely to be a simple relation.}. Thus, let us consider the problem in the ``entropy representation'', which involves evolving $\hat s$ rather than $\rho$.

We now take the equation of state to be $\varepsilon=\varepsilon(n,s)$. The evolution problem is then given by
\eqref{baryons} and \eqref{momentum}, as before, together with \eqref{entrops}, which provides $\hat s = sW$. The inversion to the primitive variables  remains a one-dimensional root-finding exercise. As in the cold fluid case, we start by guessing $n=\bar n$. The ratio of the evolved variables $\hat s$ and $\hat n$ then gives the specific entropy so the entropy density corresponding to our guess is
\be
s = \bar n ( \hat s /\hat n)  \ .
\label{sbar}
\ee
Thus we have the two parameters we need to work out $v^i$ and $W$ from the the evolved momentum $S^i$, exactly as before. Finally, we arrive at $\hat n=\bar n W$ which we compare to the evolved value and iterate until the solution is found.

The introduction of additional fluxes, like $\hat J^i$ and $Q^i$, adds steps to the inversion procedure, but it remains (at least in principle) a nonlinear root-finding problem that is qualitatively similar to the single fluid case. As the dimension of the root-finding procedure increases, it becomes more sensitive to the initial guess, more computationally expensive (usually as the square of the dimension), and less robust. In order to outline the procedure, it is useful to consider three problems of increasing complexity.

Let us first assume that the entropy remains locked to the baryons, as in the hot model discussed earlier. There is no heat flux, but the introduction of the charge current as a dynamical quantity means that we need to evolve $\hat n_\e$ and $\hat J^i$. Nevertheless, the inversion to the primitive variables remains a one-dimensional root search. We can initiate this as before; given a guessed value $n = \bar n$ we get the entropy density from \eqref{sbar} and we also have
\be
n_\e = \bar n ( \hat n_\e /\hat n)  \ .
\label{nebar}
\ee
This provides all information required to use the equation of state to evaluate $\varepsilon$, $p$ and the electron chemical potential $\mu_\e$.  Given this information we can solve the (linear in drift velocities) system 
\be
S^2 \approx (p+\varepsilon) \left[ (p+\varepsilon) W^4 \hat v^2 - {2\mu_\e \over e} \hat \sigma W \right] \ ,
\ee
\be
S_\e^2 \approx \mu_\e^2 \left( W^2 \hat v^2 - {2W\over en_\e} \hat \sigma\right) \ ,
\ee
to obtain $W$ and $\hat \sigma$. This leads to an updated value for the number density $n=\hat n /W$ which replaces our guessed value. Iteration of the procedure leads to a consistent solution which can be used to invert $S^i$ and $S_\e^i$ to get $\hat v^i$ and $\hat J^i$.

When we introduce the heat flux, the problem becomes one level more complicated. The evolution now provides
\be
\hat s = sW\left( 1 + {1\over sT} \hat v_j Q^j\right)  \ ,
\ee
and
\be
S_\s^i = T{\hat s\over s}\hat v^i + {1\over sW} Q^i  \ .
\ee
That is, we need both $T$ and $\hat v_j Q^j$ in order to invert the relation for $s$. The upshot of this is that we need a two-dimensional root search. If we guess both $n=\bar n$ and $s=\bar s$, then we can work out the corresponding value for the temperature $T$ from the thermodynamics. Once we have this information, we can solve the system provided by $\hat s$ together with
\be
S^2    \approx (p+\varepsilon) W^2 \left[ (p+\varepsilon) W^2 \hat v^2 +2 (1-2W^2 \hat v^2)\hat v_j Q^j \right]  \ ,
\ee
and
\be
S_\s^2 \approx  {\hat s T\over s} \left[  \left( {\hat sT\over s} \right) \hat v^2 +  {2\over sW} \hat v_j Q^j \right]  \ ,
\ee
to get $W$, $\hat v_j Q^j$ and a new value for $s$. Once we have the Lorentz factor we also have a new value for $n$ and we can iterate. After finding a consistent solution, we invert the expressions for the evolved fluxes to get $\hat v^i$ and $Q^i$.

The general case with three fluxes does not involve any additional complications. It remains a two-dimensional root search, as in the case with heat flux. We need to solve a system of four equations following from $\hat s$, $S^2$, $S_\e^2$ and $S_\s^2$ to determine $s$, $W$,  $\hat \sigma$ and $\hat v_j Q^j$. This gives new values for $n$ and $s$ for which we iterate. Once we have an iterated solution, we solve the coupled system for the three fluxes to get $\hat v^i$, $J^i$ and $Q^i$.

\section{Going further: The entrainment}

In the specific models considered in the previous section the entrainment effect was not included. However, there are a number of cases where entrainment may be crucial, such as causal heat propagation~\cite{heat2}. Models including entrainment will pose some novel problems for numerical simulations.

\subsection{Balance law form}

The general multi-fluid formulation  gives equations of motion that, on writing them in a 3+1 foliation point-of-view, will appear in the quasilinear form
\begin{equation}
  \partial_t {\bf U} + A^{(i)} \partial_i {\bf U} = {\bf S} \ .
\end{equation}
When the matrices $A^{(i)}$ can be written as Jacobians $\partial {\bf F}^{(i)} / \partial {\bf U}$ then the quasilinear form can be written as balance laws,
\begin{equation}
  \partial_t {\bf U} + \partial_i {\bf F}^{(i)} {\bf U} = {\bf S} \ .
\end{equation}

This makes a crucial difference when considering discontinuous solutions, particularly shocks, which are expected to appear generically in nonlinear hydrodynamics, and whose behaviour is important in astrophysical situations such as neutron star mergers or supernovae. The speed $V_S$ of a discontinuity connecting state ${\bf U}_L$ to state ${\bf U}_R$ must satisfy
\begin{equation}
  V_S \left( {\bf U}_R - {\bf U}_L \right) = \int_{{\bf U}_L}^{{\bf U}_R} A \, \text{d} {\bf U} \ .
\end{equation}
When the matrix $A$ can be written as a Jacobian this gives the standard Rankine-Hugoniot conditions. In the general non-conservative quasilinear case, however, the more general theory of~\cite{dal1995definition} is required, where the shock speed directly depends on the path in state space connecting ${\bf U}_{L, R}$, and there is no \emph{a priori} reason for choosing one path over another. Additional physical input will be needed to fix the shock speed.

Even after choosing a path, problems remain in performing a numerical simulation. Whilst a number of \emph{path-conservative} methods have been constructed to deal with the resulting non-conservative equations (once a path has been chosen), there are cases (see~\cite{Castro2008} and~\cite{Abgrall2010}) where the numerical scheme does not converge to the expected solution and different numerical methods do not agree.

It is important, therefore, to know when our general framework allows us to write the foliation equations in balance law form. We first note that the single fluid momentum equation
\begin{equation}
  2 n^b \nabla_{[b} \mu_{a]} = 0 \ , 
\end{equation}
can be re-written in the form
\begin{equation}
  \nabla_b \left[ n^b \mu_a - \delta^b_a \left( n^c \mu_c + \Lambda \right) \right] = 0 \ .
\end{equation}
Projecting this into the foliation clearly gives the balance law form expected for single fluid hydrodynamics. With this in mind we consider when the general form for the momentum equation for a single species, equation~\eqref{intmom}, can be written in the ``balance law'' form
\begin{equation}
  \nabla_b \left[ n_\x^b \mu^\x_a - \delta^b_a \left( n_\x^c \mu^\x_c + E_\x \right) + D_{\x\phantom{b}a}^b \right] = S^\x_a \ , 
\end{equation}
where the ``source'' $S^\x_a$ contains no derivatives of fluid variables. By considering $E_\x$ and $D_{\x\phantom{b}a}^b$ to be functions of $n_{\x}^a$ only, we can see that this matches equation~\eqref{intmom} \emph{only} if
\begin{equation}
  S^\x_a = R^\x_{b} + \left[ \frac{\partial D_{\x\phantom{c}b}^c}{\partial n_{\x}^{a}} - \delta^c_b \left( \mu^{\x}_{a} + \frac{\partial E_{\x}}{\partial n_{\x}^{a}} \right) \right] \nabla_c n_{\x}^{a} +  \sum_{\y\ne\x} \left[ \frac{\partial D_{\x\phantom{c}a}^c}{\partial n_{\y}^{a}} - \delta^c_b  \frac{\partial E_{\x}}{\partial n_{\y}^{a}} \right] \nabla_c n_{\y}^{a} \ .
\end{equation}
For the source term to contain no derivatives of fluid variables requires that
\begin{equation}
  \frac{\partial \mu^\x_a}{\partial n^\y_d} \equiv 0 \ , 
\end{equation}
which is precisely when there is no entrainment. We therefore expect that it will not be possible to write all equations for models including entrainment in balance law form~\footnote{The equations for total energy and momentum conservation will obviously remain of the required form as long as the system is closed.}, meaning the complexities of non-conservative equations of motion and path-conservative numerical methods will be needed.

\subsection{Inferring the primitive variables}

In section~VF the reconstruction of the primitive variables from the evolved variables was more complex than the single fluid problem, but remained a relatively straightforward root-finding problem.

When entrainment is included the couplings between different species introduced by the equation of state become more complex. In particular, computing any entrained conjugate momentum requires knowing the number density and velocities of all relevant species. In the most general case where all species are entrained, it will be necessary to solve simultaneously for all species number densities and velocities, which increases the dimensionality of the root-finding problem substantially.

The steps required were essentially laid out in~\cite{Hawke2013}. In summary, we would guess the number densities of all species. Given the evolved variables in the individual momentum equations, which are proportional to the conjugate momenta, we can use the definition of the conjugate momenta to solve a linear system for the (spatial components) of the species velocities. From this and the evolved variables from the individual continuity equations we can get the number densities. This gives a root-finding problem whose size corresponds to the number of species, which is likely to be costly and numerically sensitive to, for example, the choice of initial guess. As an example, for a general three fluid model the two dimensional root find using algebraic relations in section VF would be replaced by a three dimensional root find involving a linear system solve at each stage, which will likely at least double the computational cost.

\section{Summary and outlook}

We have considered the  general resistive multi-fluid framework discussed in~\cite{fibrate} from a 3+1 space-time foliation point-of-view. With future numerical simulations in mind we paid particular attention to issues relating to the use of multi-parameter equations of state and the associated inversion from evolved to primitive variables. We highlighted numerical issues that arise for systems with relative flows and the entrainment coupling. One important technical issue that remains to be resolved arises from the fact that the general multi-fluid problem cannot be cast in flux-conservative form, and we touched upon possible challenges this may lead to. As an example of the new formulation, we focussed on a three-component system relevant for hot neutron stars. We assumed the baryons (neutrons and protons) move together, but let heat and electrons exhibit relative flow. This reduces the problem to three momentum equations; overall energy-momentum conservation, a generalised Ohm's law and a heat equation. Our results provide a hierarchy of increasingly complex models for this system and prepare the ground for more detailed  state-of-the-art simulations of relevant relativistic scenarios.

The  natural next step is to carry out numerical simulations to test the relevance of the new features accounted for in our model. Work in this direction is in progress. For example, we consider different aspects of resistive two-component plasmas in~\cite{kiki1}. At the moment, the numerical work is very much at the development stage. While we make progress on the computational side, we also need to develop the formal theory further. In particular, we need to include radiative aspects in order to be able to account for neutrino emission if we want to accurately model hot systems. The models we developed in this paper may contain trapped neutrinos (forming part of the entropy component) but we did not account for possible radiative fluxes. However, the strategy for adding these aspects is, at least in principle, clear (see, e.g., \cite{Cardall2013} or \cite{Shibata2014}). Similarly, the general framework is readily extended to include the elastic neutron star crust, which will be relevant for mature (cold) systems~\cite{elastic}. Once the model is extended in these directions we will have a flexible theoretical framework which will allow us to model the nonlinear dynamics of neutron stars at all stages of evolution, from birth to maturity (and perhaps, as the magnetic field decays, obscurity).

\acknowledgments

NA, IH and KD gratefully acknowledge support from the STFC.

\section*{Appendix: The electromagnetic field}

For completeness, we provide the relevant evolution equations for the electromagnetic degrees of freedom in this Appendix. There are different approaches to this part of problem. The electromagnetic dynamics is fully specified in terms of  the vector potential $A^a$, but it may be more intuitive to work with the electric and magnetic fields, $E^a$ and $B^a$. Our formulation of the fluid part of the problem is non-committal in this respect, but it is worth noting that we need to evaluate the vector potential whenever we want to account for particle reactions. This inevitably involves electromagnetic gauge issues~\cite{variate} which suggests that a formulation like that discussed in~\cite{baum} (which involves $E^a$ and $A^a$) may be natural.

Postponing a deeper discussion of this issue for the future, let us assume that we work with the electric and magnetic fields.
 In the 3+1 decomposition, where the observer is associated with $N^a$, we then have
\be
F_{ab} =  2 N_{[a} E_{b]}  + \epsilon_{abcd}N^c B^d \ ,
\label{Faraday_appendix}\ee
That is, the electric and magnetic fields measured in the Eulerian frame are
\be
E_a = -  N^b F_{ba} \ ,
\ee
and
\be
B_a = - N^b \left( {1 \over 2} \epsilon_{abcd}F^{cd}\right) \ .
\ee
The  fields are both orthogonal to $N^a$, so
each  has three components, just as in non-relativistic physics.

It is useful to relate the fields to those associated with the frame used in~\cite{fibrate}, where we had
(using lowercase letters represent the  fields measured in the fluid frame associated with $u^a$)
\be
F_{ab} =  2 u_{[a} e_{b]}  + \epsilon_{abcd}u^c b^d \ ,
\ee
We need 
\begin{multline}
e_a = - u^b F_{ba} = - W(N^b +v^b) F_{ba} = W\left[ E_a +N_a (\hat v^b E_b)\right] - W \hat v^b \epsilon_{bacd} N^c B^d \\
= W \left[E_a + N_a (\hat v^b E_b) + \epsilon_{abc}\hat v^b B^c \right] \ ,
\end{multline}
and
\begin{multline}
b_a = - u^b \left( {1 \over 2} \epsilon_{abcd} F^{cd}\right) = - W(N^b +\hat v^b) \left( {1 \over 2} \epsilon_{abcd} F^{cd}\right) \\
=  W \left[ B_a +N_a ( \hat v^b B_b) + \epsilon_{abc} \hat v^b E^c  \right] \ .
\end{multline}

The electromagnetic contribution to the stress-energy tensor is
\be
T^\mathrm{EM}_{ab} = {1\over \mu_0} \left[ g^{cd} F_{ac} F_{bd} - {1\over 4} g_{ab} (F_{cd} F^{cd} ) \right] \ .
\ee
In terms of the the fields (measured by the Eulerian observer) we have
\be
T^\mathrm{EM}_{ab} = E^2 N_a N_b + E_a E_b + \gamma_{ab} B^2 - B_a B_b + 2N_{(a}\epsilon_{b)dh} e^d B^h - {1\over 2} g_{ab} \left( B^2 - E^2\right) \ ,
\ee
from which \eqref{magrho}--\eqref{magsij} follow.

Rather that working with the divergence of the total stress-energy tensor for the system we can isolate  the electromagnetic contribution. The right-hand side of the matter equations then have additional terms which follow from the Lorentz force;
\be
f_\mathrm{L}^a = - j_a F^{ab} = N^b (\hat J^a E_a) + ( \hat \sigma E^b + \epsilon^{bad} \hat J_a B_d ) \ ,
\ee
where the charge current
\be
j^a = \hat \sigma N^a + \hat J^a \ ,
\ee
was discussed in the main text of the paper.

Finally,  we need Maxwell's equations. First of all,
\be
\nabla_a F^{ba} = \mu_0 j^b \ ,
\ee
leads to
\be
\gamma^{ab} \nabla_b E_a = \mu_0 \hat \sigma + \epsilon^{abc}\left( \nabla_a N_b\right) B_c \ , 
\ee
or
\be
\gamma^{b}_a\nabla_b E^a - \mu_0 \hat \sigma = - \epsilon^{abc}K_{ab} B_c = 0 \ ,
\ee
since $K_{ab}$ is symmetric. That is, we have
\be
D_i E^i = \mu_0 \hat \sigma \ .
\ee

We also get
\be
\gamma_{ab} N^c\nabla_c E^b - \epsilon_{abc}\nabla^b B^c + \mu_0 \hat J_a =  E^b \nabla_b N_a - E_a \nabla_b N^b  + \epsilon_{abc}( N^d \nabla_d {N}^b) B^c \ ,
\ee
or
\be
\gamma_{ab} N^c\nabla_c E^b - \epsilon_{abc}\nabla^b B^c + \mu_0 \hat J_a =  - E^b K_{ba} + E_a K  + \epsilon_{abc}( N^d \nabla_d {N}^b) B^c \ ,
\ee
and we end up with
\be
\left( \partial_t - \mathcal L_\beta\right) E^i - \epsilon^{ijk} D_j (\alpha B_k) + \alpha \mu_0 J^i = \alpha K e^i \ .
\ee

The second pair of Maxwell equations follow from
\be
\nabla_{[a}F_{bc]} = 0 \ ,
\ee
which leads to
\be
\gamma^{ab}\nabla_b B_a = -  \epsilon^{abc} E_a \nabla_b N_c \ ,
\ee
or
\be
\gamma^{b}_a\nabla_b B^a =  \epsilon^{abc} E_a K_{bc} = 0 \ ,
\ee
So we have
\be
D_i B^i = 0 \ .
\ee

Finally,
\be
\gamma_{ab}N^c\nabla_c {B}^b +  \epsilon_{abc}\nabla^b E^c   = - \epsilon_{abc}(N^d \nabla_d N^b) E^c +   B^b \nabla_b N_a - B_a \nabla_b N^b \ ,
\ee
or
\be
\gamma_{ab}N^c\nabla_c {B}^b +  \epsilon_{abc}\nabla^b E^c   = - \epsilon_{abc}(N^d \nabla_d N^b) E^c -   B^b K_{ba} + B_a K \ .
\ee
This leads to
\be
\left( \partial_t - \mathcal L_\beta\right) B^i + \epsilon^{ijk} D_j (\alpha B_k) = \alpha K B^i \ .
\ee

\end{document}